\documentclass[12pt]{iopart}
\usepackage{iopams}
\usepackage{ifpdf}
\ifpdf
  \usepackage[pdftex]{graphicx,hyperref}   
\else
  \usepackage[dvips]{graphicx, hyperref}		
\fi

\newcommand{\des}{\Psi^{}}
\newcommand{\cre}{\Psi^\dagger}

\newcommand{\spin}{\sigma}

\newcommand{\gap}{\Delta_{\spin \spin'}}

\newcommand{\ls}{\left(}
\newcommand{\rs}{\right)}
\newcommand{\lh}{\left[}
\newcommand{\rh}{\right]}

\newcommand{\Ham}{H}
\newcommand{\om}{\omega}

\newcommand{\mb}[1]{\mathbf{#1}}
\newcommand{\mt}[1]{\mathrm{#1}}

\begin{document}

\title{Coexistence of pairing gaps in three-component Fermi gases}
\author{O.H.T. Nummi, J.J. Kinnunen and P. T\"orm\"a}
\address{Department of Applied Physics, P.O. Box 5100, 02015 Aalto University, Finland}
\ead{paivi.torma@hut.fi}
\begin{abstract}
We study a three-component superfluid Fermi gas in a spherically 
symmetric harmonic trap using the Bogoliubov-deGennes method. We 
predict a coexistence phase in which two pairing field order parameters
are simultaneously nonzero, in stark contrast to studies performed
for trapped gases using local density approximation. We also discuss the 
role of atom number conservation in the context of a homogeneous system.
\end{abstract}


\submitto{\NJP}

\maketitle

\section{Introduction}

Multicomponent ultracold Fermi gases allow the study of several interesting questions in 
many-body quantum physics.
In particular, understanding three-component pairing can reveal some properties 
of multi- or two-component pairing. In three-component systems the pairing energy does 
not only compete with temperature effects or Fermi surface mismatch energy 
but also with other pairing gaps. As in imbalanced two-component systems, 
non-BCS pairing mechanisms such as Larkin-Ovchinnikov-Fulde-Ferrel (LOFF) 
\cite{FF, LO}, breached pairing (BP) \cite{BP} and phase separation phases
are expected.

Ultracold Fermi gases have opened up a way to explore multi-component 
gases experimentally. Recently, a degenerate three-component gas was successfully created 
\cite{Ottenstein2008a,Huckans2009a}. The stability of three-component gases is 
hindered by the three-body recombination, reducing the pairing in the 
gas and the lifetime of the sample \cite{Williams2009a, Wenz2009a, Lompe2010a}.
However, there are ways to stabilize the gas against such losses
using for example optical lattices. Optical lattices are interesting
also due to the rich phase diagram: theoretical investigations have found
that color superconductivity competes with normal phase and 
formation of trions \cite{Rapp2007a, Capponi2008a, Inaba2009a, Azaria2009a, Kantian2009a, Klingschat2010a,Luscher2009a}.

Both the SU(3) symmetric model, in which the different
components have identical properties, and the non-SU(3) symmetric case have been widely 
studied~\cite{Modawi1998a, Honercamp2004a, Honercamp2004b,Paananen2006a,He2006a,Cherng2007a,Catelani2008a,Bedaque2009a,Chiacchiera2010a,Ozawa2010a}. 
The particularly important special case of SU(3) symmetry can be realized using alkaline earth atoms~\cite{Tey2010a} or in optical lattices. However, 
with alkaline earth atoms the numbers of atoms in different components are 
not well defined, and only the
total number of atoms is conserved. In contrast, the hyperfine energy spacing
in alkaline atoms stabilizes the atom numbers, making the atom number of each
component separately conserved. This is the case of most two-component Fermi gas
experiments and a natural extension of these studies is a non-SU(3) symmetric case in
which the atom numbers in all three components are fixed.

Here we study such a three-component system with fixed atom numbers in all three components
in a spherically symmetric harmonic trap using the Bogoliubov-deGennes (BdG) equations. 
We study the coexistence of the pairing gaps in these systems and discuss the scaling of the system 
size up to the thermodynamical limit.

In section~\ref{sec:setup} we give an overview of the three-component system 
and the corresponding mean-field theory. 
In the next section~\ref{sec:homogeneous} we consider the BCS-type mean-field theory in homogeneous
space and describe the effect of boundary conditions on the stability of different phases. In section~\ref{sec:BdG} we consider the effects of trapping potential using the BdG method. In section~\ref{sec:results} we show the main results obtained from the BdG method, especially regarding the coexistence of
two pairing gaps. We conclude by discussion in section~\ref{sec:conclusions}.

\section{The system setup}
\label{sec:setup}
The general mean-field Hamiltonian for a three-component system in the contact interaction potential approximation is (up to a constant) 
\begin{eqnarray}
\label{Ham1}
H_{\mt{MF}} &= \sum_{\spin = 1,2,3} \int d^3 r\,  \cre_\spin(\mb r)\lh -\frac{\hbar^2 \nabla^2}{2m_\spin} + V_\spin (\mb r) - \mu^{}_\spin + W_\spin(\mb r) \rh \des_\spin(\mb r)  \\
&+ \frac{1}{2}\sum_{\spin \neq \spin'} \int d^3 r\, \Delta_{\spin \spin'}(\mb r) \cre_\spin(\mb r) \cre_{\spin'}(\mb r)  + h.c., 
\end{eqnarray}
where the first term of the Hamiltonian includes contributions from the kinetic energy, the external 
trapping potential $V_\spin(\mb r)$ (that can depend on the component $|\spin \rangle$), and the chemical potentials 
$\mu^{}_\spin$, respectively. Interactions are described by the two-body scattering T-matrix. 
In the contact interaction potential approximation it can be written as 
\begin{equation}
 U_{\spin \spin'}(\mb r, \mb r') = \frac{4\pi \hbar^2 a_{\spin \spin'}}{m_r} \delta(\mb r - \mb r'),
\end{equation}
where $a_{\spin \spin'}$ is the scattering length between atoms in hyperfine states 
$|\spin \rangle$ and $| \spin' \rangle$, and $m_r = 2 m_\spin m_{\spin'}/(m_\spin + m_{\spin'})$ is 
twice the reduced mass. The Hartree fields are denoted by 
$W_\spin(\mb r) = \sum_{\spin \neq \spin'} U_{\spin \spin'}(\mb r)\, n_{\spin'}(\mb r)$ and 
the densities are $n_{\spin}(\mb r) = \langle \Psi^\dagger_\spin(\mb r) \Psi^{}_\spin(\mb r) \rangle$.
The pairing (mean-)field 
$\Delta_{\spin \spin'} (\mb r) = \tilde U_{\spin \spin'}(\mb r) \langle \Psi_\spin(\mb r) \Psi_{\spin'}(\mb r) \rangle$ includes a renormalized
interaction $\tilde U_{\spin \spin'}(\mb r)$ that is used to remove the ultraviolet divergence following the standard procedure (see below).
In our model we neglect the possibility of three-body bound states and
other three-body effects that can affect the lifetime of the 
gas~\cite{Wenz2009a}. 

A three-component system has three possible pairing fields corresponding to the three
interaction channels $U_{12}$, $U_{13}$, $U_{23}$, and these can be 
combined into a total pairing field vector 
$\mbox{\boldmath$\Delta$} = [ \Delta_{23}, - \Delta_{13}, \Delta_{12}]^\mt T$. 
Identical properties make the system SU(3)-symmetric and the pairing vector can be 
reduced to a single gap by a simple unitary transformation. The orientation
of the pairing vector corresponds to a choice of the global 
gauge~\cite{Modawi1998a, Honercamp2004a, Honercamp2004b,Paananen2006a,Ozawa2010a}, and the simplest choice 
is the one where only one of the pairing gaps, say $\Delta_{12}$, is nonzero.
This of course makes atoms in component $|3\rangle$ effectively 
noninteracting. Indeed,
in an SU(3)-symmetric case, there always exists a gapless branch describing  
unpaired atoms.

In our study one interaction is always suppressed (we choose $U_{13} = 0$). 
This is the case for example in $^6$Li where Feshbach resonances 
between the three lowest hyperfine states $|1\rangle-|2\rangle$ 
(at $B = 834\,\mathrm{G}$) and $|2\rangle-|3\rangle$ (at $B=811\,\mathrm{G}$) 
lie close to each other, while the resonance $|1\rangle-|3\rangle$ 
(nearest one at $B = 690\,\mathrm{G}$) is sufficiently far 
away~\cite{Bartenstein2005a}. 
Similar behavior occurs in 
$^{40}$K~\cite{Regal2004a}, where the richness of the 
hyperfine level structure allows even more freedom in choosing the 
suitable interaction strengths. Moreover, mixtures of $^6$Li and $^{40}$K offer
interesting possibilities~\cite{Spiegelhalder2009a,Spiegelhalder2010a}. Also we do not consider here interactions
between $|1\rangle$ and $|3\rangle$ induced by the component 
$|2\rangle$~\cite{Martikainen2009a}.
Thus, neglecting the $|1\rangle-|3\rangle$ interaction channel altogether, 
the symmetry is broken at least to SU(2)$\times$SU(1).
Analogously to the SU(3) symmetric case, the total pairing field
is now a two-dimensional vector 
$\mbox{\boldmath$\Delta$} = [ \Delta_{12}, \Delta_{23}]^\mt T$ and in the
symmetric case, where components $|1\rangle$ and $|3\rangle$ are identical,
it is preserved under spin rotations of the 
hyperfine states $|1\rangle$ and $|3\rangle$.
This symmetry implies that the ground state is degenerate with respect
to the orientation of the total pairing field vector. 
However, the degeneracy is lifted by changing masses, chemical potentials
or interaction strengths and, as we will soon show, also by imposing
boundary conditions such as fixing the number of atoms in different components.

Boundary conditions, such as fixed particle numbers or fixed chemical
potentials, manifest themselves in different ways in atomic gases. While the
total particle numbers are, in practice, fixed, the local densities are not 
as the particles are allowed to move around in the trap. Hence,
from the local density approximation (LDA) point of view, locally the relevant
boundary condition appears to be a fixed chemical potential. 
However, globally the relevant boundary condition is the fixed particle number, and in the BdG method we indeed fix the mean particle number.
Below we will also discuss how 
the two pictures merge in the limit of large system size $N \rightarrow \infty$.

\section{Homogeneous system}
\label{sec:homogeneous}

For a homogeneous system the densities and gaps lose their spatial dependence. 
This corresponds to the usual BCS-approximation in which pairing can only occur between 
atoms with opposite momenta $|\mb k, \spin \rangle$ and $|-\mb k, \spin' \rangle$. 
The mean-field Hamiltonian can be written in matrix form as
\begin{equation}
\fl
\Ham_{\mt{MF}} = \sum_{\mb k}
\left(\begin{array}{c}
c^{\dagger}_{1 \mb k}  \\
c_{2 -\mb k}  \\
c^{\dagger}_{3 \mb k}  \\
\end{array}\right)^\mt{T}
\left(\begin{array}{ccc}
\xi^{}_{1 \mb k}  & \Delta_{12} & 0  \\
\Delta_{12} & -\xi^{}_{2 -\mb k} & \Delta_{23}  \\
0 & \Delta_{23} & \xi^{}_{3 \mb k}  \\
\end{array}\right)
\left(\begin{array}{c}
c^{}_{1 \mb k} \\
c^{\dagger}_{2 -\mb k} \\
c^{}_{3 \mb k}  \\
\end{array}\right)
+ C,
\end{equation} 
where $C$ is constant and the single-particle dispersion is 
$\xi^{}_{\spin \mb k} = \frac{\hbar^2 k^2}{2m_\spin}-\mu^{}_\spin$. We have here 
neglected the Hartree fields since at the level of our approximation they provide 
only a constant energy shift in a homogeneous system.
The standard approach calls for diagonalizing this using the Bogoliubov transformation, 
and iteratively solving for the pairing fields $\Delta_{12}$ and $\Delta_{23}$. In order 
to satisfy the fixed mean atom number boundary condition, the iteration must adjust the 
chemical potentials in a self-consistent manner as well. Notice that the 
inherent atom number fluctuations implied by the mean-field 
theory play no role here as long as the typical fluctuations (scaling as 
$\sqrt{N}$) are much smaller than the atom numbers in different species (scaling as $N$).

\begin{figure}
  \centering
  \includegraphics[width=0.45\textwidth]{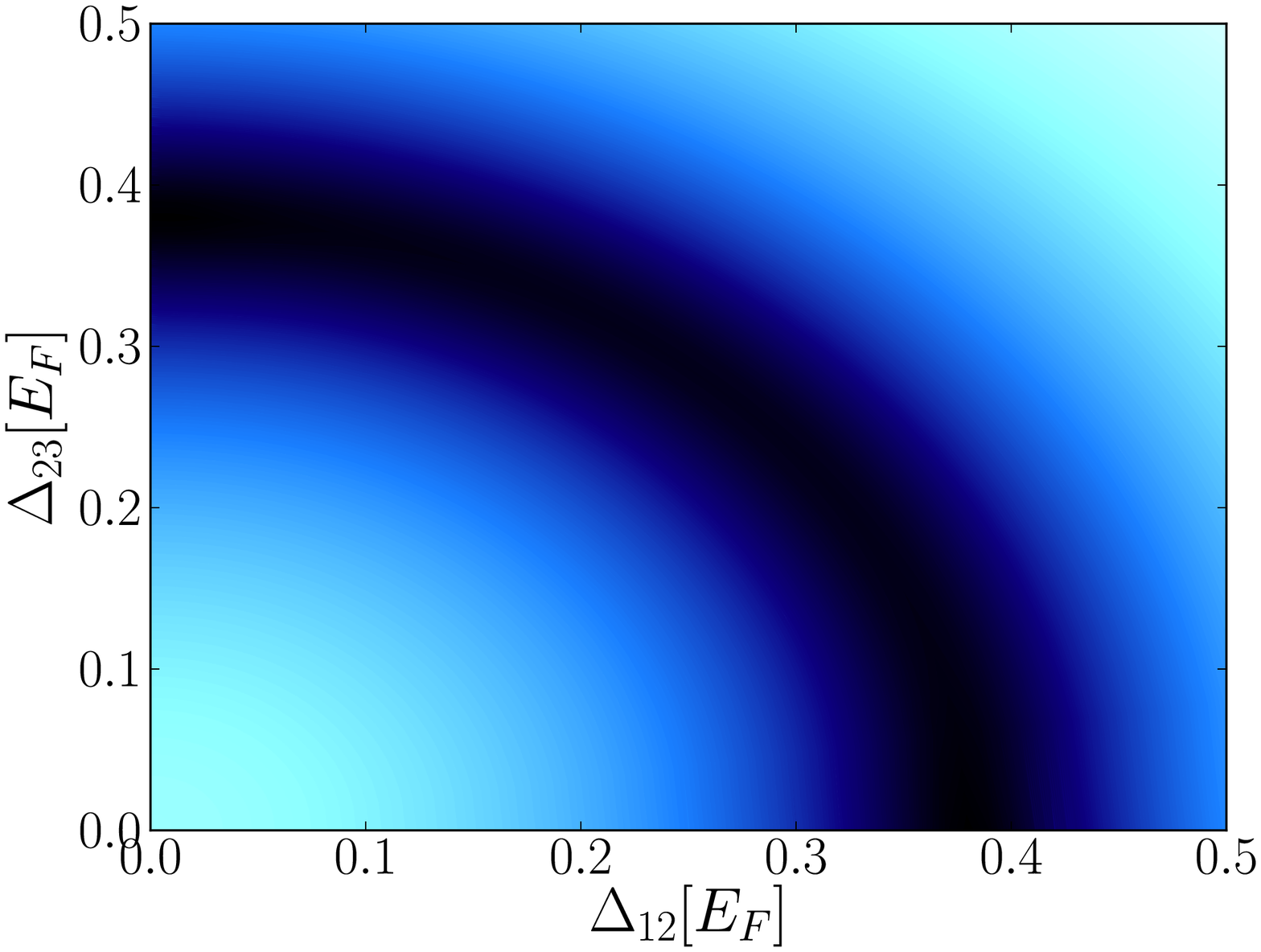}
  \includegraphics[width=0.45\textwidth]{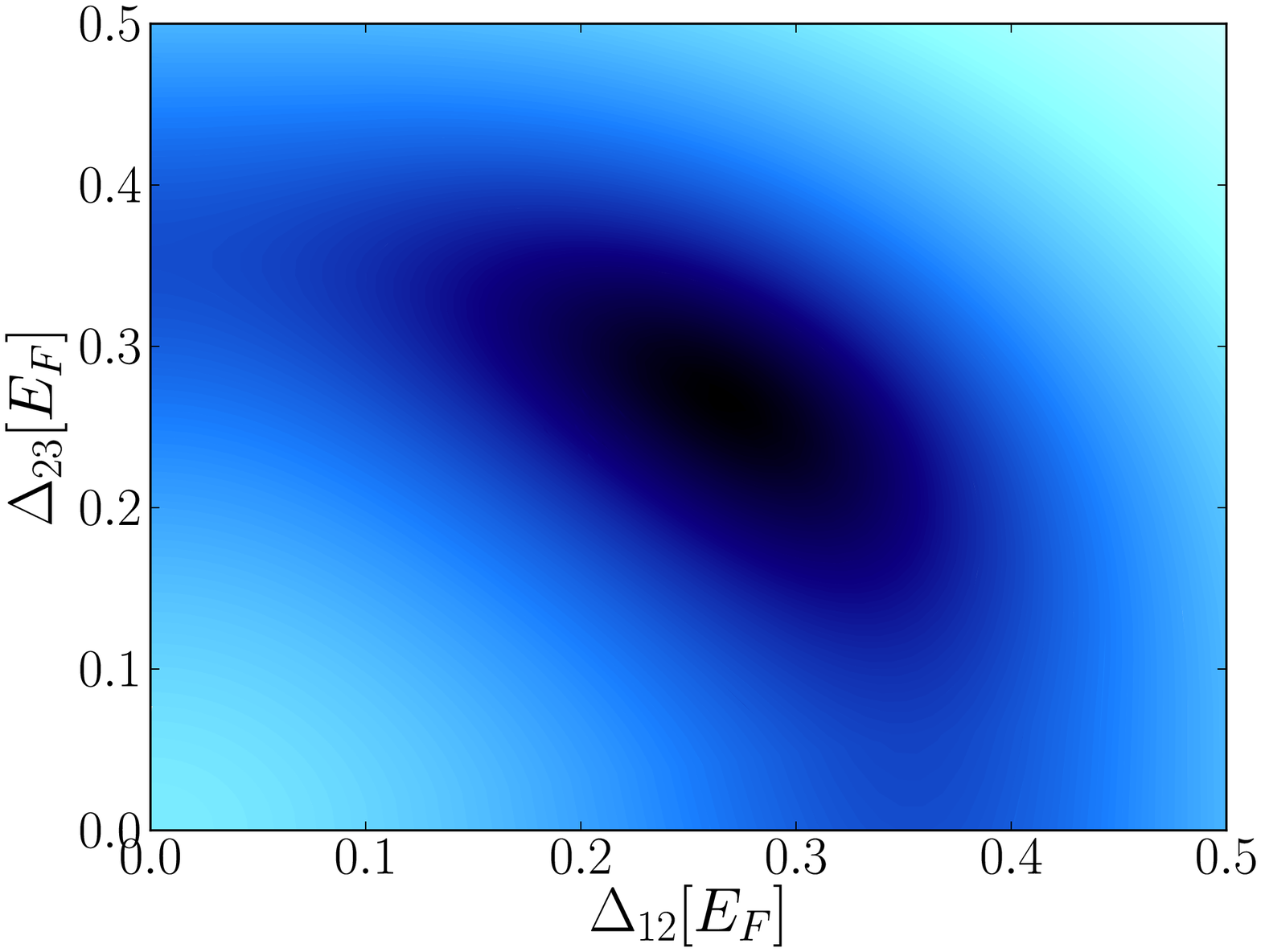}
  \includegraphics[width=0.45\textwidth]{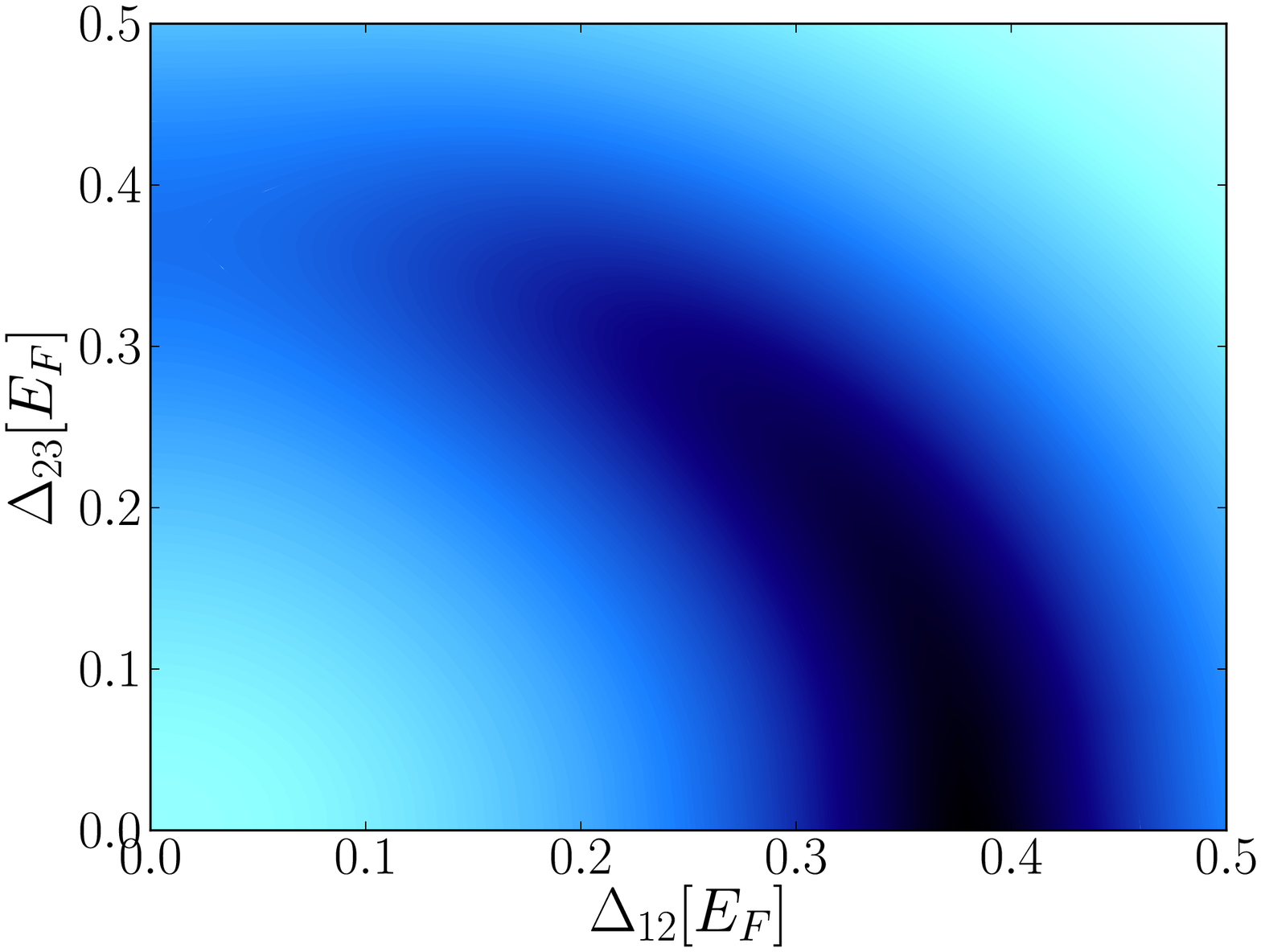}
  \includegraphics[width=0.45\textwidth]{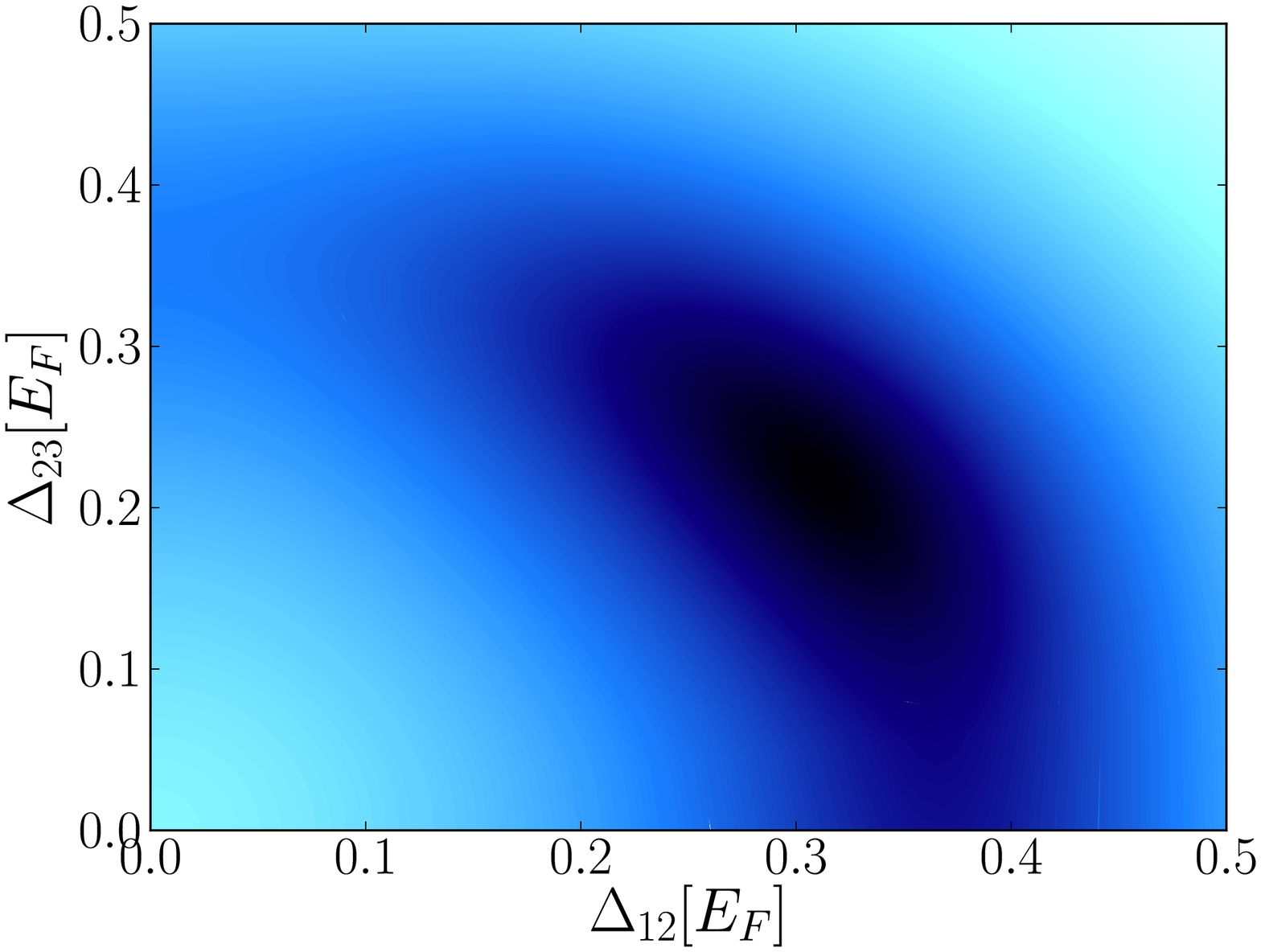}
  \caption{Logarithmic energy landscape with constant particle numbers (a) $N_1 = N_3 = 1.0N_2$ (b) $N_1 = N_3 = 1.2N_2$ (c) $N_1 = N_2, N_3 = 1.1N_2$ (d) $N_1 = 1.1 N_2,  N_3 = 1.2N_2$. The interaction strengths are equal $(k_\mathrm{F} a_{12})^{-1} = (k_\mathrm{F} a_{23})^{-1} = -0.5$, ($k_\mathrm{F} = (6\pi^2 n_2)^{1/3}$).}
\label{fig:energies}
\end{figure}

As discussed above, in the symmetric case ($\mu^{}_1 = \mu^{}_3$, 
$m_1 = m_3$, $U_{12} = U_{23}$), 
the Hamiltonian has SU(2)$\times$SU(1) symmetry and all the pairing fields 
with $\Delta_{12}^2 + \Delta_{23}^2$ constant yield the same total energy.
The ground state is thus degenerate. However, different orientations of the 
pairing field vector yield different atom numbers in components $|1\rangle$ 
and $|3\rangle$.
Thus, fixing the numbers of atoms $N_1$ and $N_3$ breaks the degeneracy
and a well-defined energy minimum is found.
Figure~\ref{fig:energies} shows typical energy landscapes 
$\langle \Ham_{\mt{MF}} + \sum_\spin \mu^{}_\spin N_\spin \rangle$ as a 
function of the pairing fields $\Delta_{12}$ and $\Delta_{23}$
for equal interaction strengths $U_{12} = U_{23}$. The Fermi momentum $k_\mathrm{F}$
here and throughout this work is defined as the Fermi momentum of the component
$|2\rangle$, $k_\mathrm{F} = \sqrt{2m E_\mathrm{F}^{\spin = 2}}/\hbar$, where $E_\mathrm{F}^\spin$ is
the Fermi energy of the component $|\spin\rangle$.
The energies have been calculated for fixed atom numbers: the
chemical potentials $\mu^{}_\spin$ are solved for every point $(\Delta_{12}, \Delta_{23})$ 
so that the atom number constraints are satisfied. The figures show clearly
how the ground state becomes non-degenerate and realizes itself
in a particular combination of pairing fields. In the special case where
there is equal number of atoms in all three components, $N_1 = N_2 = N_3$, 
the ground state is still degenerate. However, if the number of atoms 
in component $|2\rangle$ is changed (keeping $N_1=N_3$ but $N_2 \neq N_1$), 
the degeneracy is broken and a non-degenerate energy minimum appears. 
This is in stark contrast to the case where the chemical potentials are 
kept constant and the atom numbers are allowed to vary. In such a case the 
ground state remains degenerate as long as the chemical potentials for 
components $|1\rangle$ and $|3\rangle$ are equal, $\mu_1 = \mu_3$.

\begin{figure}
  \centering
  \includegraphics[width=0.7\textwidth]{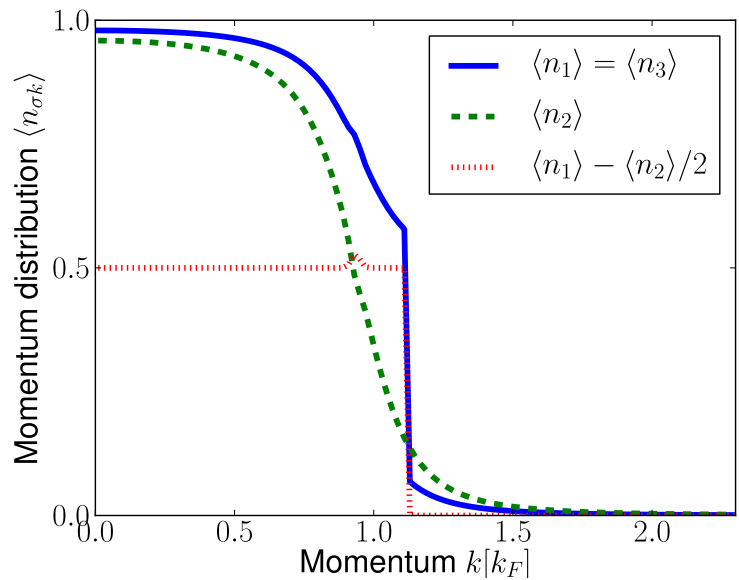}
  \caption{The single-particle occupation numbers for $N_1 = N_3 = 1.2N_2$ and equal pairing fields
$\Delta_{12} = \Delta_{23} = 0.25 E_\mathrm{F}$. All atoms in the hyperfine state $|2\rangle$ are 
paired but part of the atoms in components $|1\rangle$ and $|3\rangle$ are unpaired. The unpaired
atoms form a well-defined Fermi sphere resulting in a step at the Fermi surface.}
\label{fig:free_momentum}
\end{figure}

In the case of a number mismatch or difference in the interaction strengths of
the components $|1\rangle$ and $|3\rangle$, the energy minimum will be shifted 
from the equal pairing case. Depending on the number of atoms in component
$|2\rangle$, the minimum appears either at the edge of the energy landscape 
(yielding either of the two pairing fields $\Delta_{12}$ or $\Delta_{23}$ zero)
or somewhere in between. This too is an important difference to the
case of fixed chemical potentials where the breaking of the symmetry
(by either changing chemical potentials or interaction strengths)
always results in either of the two pairing fields dominating and the other 
becoming zero. Thus, for fixed chemical potentials one does not observe
\emph{coexistence} of the two pairing fields $\Delta_{12}$ and $\Delta_{23}$
except possibly in the symmetric, or degenerate, case, whereas for
fixed atom numbers the coexistence phase (described by two non-vanishing
pairing field order parameters $\Delta_{12}$ and $\Delta_{23}$) is very real.

\begin{figure}
  \centering
  \includegraphics[width=0.7\textwidth]{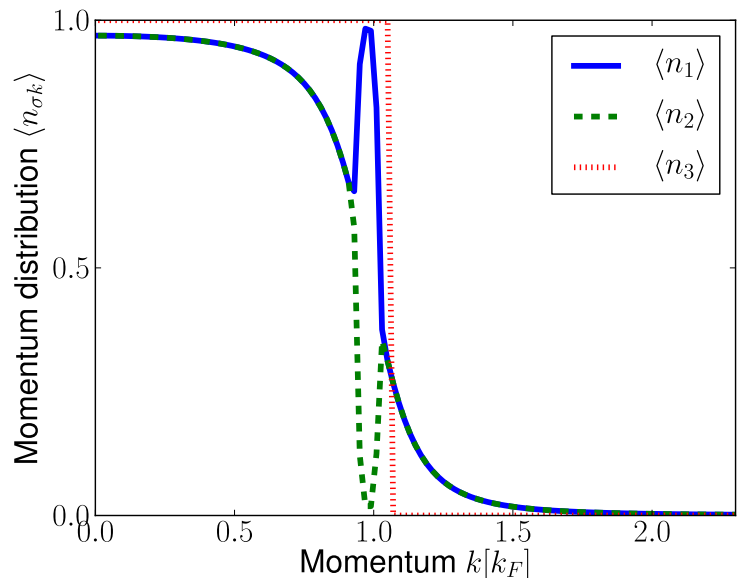}
  \caption{Single-particle occupation numbers as in Figure~\ref{fig:free_momentum} but now for projected
pairing field vector $\Delta_{23} = 0, \Delta_{12} = 0.32 E_\mathrm{F}$. All atoms in the component $|3\rangle$
are unpaired, revealing a noninteracting Fermi sea. Due to the number mismatch between components $|1\rangle$
and $|2\rangle$, excess atoms in $|1\rangle$ will form a breach.}
\label{fig:free_momentum2}
\end{figure}

The pairing scheme is revealed by the momentum distribution of each state. 
Figures~\ref{fig:free_momentum} and~\ref{fig:free_momentum2} show the momentum 
distributions of the three components for equal pairing gaps ($\Delta_{12} = \Delta_{23}$) and
for projected pairing gaps (obtained by setting $\Delta_{23} = 0$, which is always an allowed solution, 
and minimizing the energy by varying only $\Delta_{12}$.) 
In the first case, equal pairing gaps imply that there are equal numbers of 
$12$ and $23$ pairs. Since only zero-momentum Cooper pairs are considered here,
one can filter the paired atoms from the momentum distributions and determine
the momentum distribution of unpaired atoms by calculating the difference 
$\langle n_{\sigma k} \rangle - \langle n_{2 k} \rangle/2$ for $\spin = 1,3$. 
The distribution of unpaired atoms is seen to form a clear Fermi sphere, but 
with maximal occupation probability of $0.5$. In the case of projected pairing gap in
Figure~\ref{fig:free_momentum2} the pairing atoms $|1\rangle$ and $|2\rangle$ form a breach due
to a number mismatch between the two components. 

The boundary condition of fixed atom numbers for each component separately is 
natural for atomic gas experiments. However, the results for a homogeneous gas must
be approached with caution since experiments are always conducted in nonuniform trapping
potentials. Using these homogeneous system results in conjunction with local density
approximation means locally fixing the chemical potentials instead of the atom numbers. 
This discrepancy on which boundary condition to use can be solved by treating
the trapping effects explicitly using the Bogoliubov-deGennes method. 

\section{Harmonic trap -- the Bogoliubov-deGennes method}
\label{sec:BdG}

In order to consider trapped systems, we use the Bogoliubov-deGennes method
that allows the inclusion of trap effects exactly. 
The mean-field BdG method is not expected to be able to capture all relevant 
physics in the strongly interacting regime. However, in an imbalanced 
two-component system, it has been shown~\cite{Kim2011a} that,
for small polarizations and symmetric trap geometries, there is a 
good agreement between the mean-field BdG approach and real-space dynamical 
mean-field theory.
We solve the three-component mean-field system in a spherically harmonic trap $V_\spin (\mb r) = \frac{1}{2} m_\spin \omega_\spin^2 r^2$
using the eigenbasis of the 3-dimensional harmonic oscillator 
\begin{equation}
\des_\spin(\mb r) = \sum_{nlm}  R^l_{\spin n}(r) Y^{}_{lm}(\mb \Omega) c^{}_{nlm \spin},
\end{equation}
where $Y_{lm}$ are the spherical harmonics and the radial wavefunctions 
are given by
\begin{equation}
R^l_{\spin n}(r) = \sqrt{2}(m_\spin \om_{\spin})^{3/4} \sqrt{\frac{n!}{(n+l+1/2)!}} e^{-\bar r_\spin^2/2} \bar r_\spin^l L_n^{l+1/2}(\bar r_\spin^2).
\end{equation}
Here $L_n^{l+1/2}(\bar r_\spin^2)$ is the associated Laguerre polynomial and $\bar r_\spin \equiv r \sqrt{m_\spin \om_{\spin}/\hbar}$. 

The mean-field Hamiltonian separates for different $l$-quantum numbers $H = \sum_l H_l + C$, such that $\left[H,H_l\right] = 0$ and $C$ is constant. 
Introducing a finite cutoff energy $E_\mathrm{c}$ and keeping only single-particle states with energy less
than the cutoff allows writing each $H_l$ using block matrices
\begin{equation}
    \label{matrixform}
    H^l = \left( \begin{array}{c}
      \mb c^\dagger_{1l} \\
      \mb c_{2 l}\\
      \mb c^\dagger_{3 l}
    \end{array} \right)^T
    \left( \begin{array}{ccc}
       \mbox{\boldmath $\epsilon$}_{l 1} + {\bf J}^l_{12} & {\bf F}^l_{12} & {\bf 0}  \\
      \mb F^l_{12} & -\mbox{\boldmath $\epsilon$}^{}_{l 2} -\mb J^l_2 & \mb F^l_{23}  \\
      \mb 0 & \mb F^l_{23} & \mbox{\boldmath $\epsilon$}^{}_{l 3} + \mb J^l_{32} \\
    \end{array} \right)
    \left(\begin{array}{c}
      \mb c^{}_{1 l} \\
      \mb c^\dagger_{2 l}\\
      \mb c^{}_{3 l} \\
    \end{array}\right)
\end{equation}
where $\mb J^l_2 = \mb J^l_{21} + \mb J^l_{23}$.
The block matrices are defined as
\begin{equation}
\fl
    \mb J^l_{\spin \spin'} = 
    \left(\begin{array}{ccc}
      J^l_{\spin \spin' 00}& \cdots & J^l_{\spin \spin' 0N} \\
      \vdots & \ddots & \vdots \\
      J^l_{ \spin \spin' N0}  & \cdots & J^l_{ \spin \spin' NN} 
    \end{array}\right) \quad \mathrm{and} \quad
    \mb F^l_{\spin \spin'} = 
    \left(\begin{array}{ccc}
      F^l_{\spin \spin' 00}& \cdots & F^l_{\spin \spin' 0N} \\
      \vdots & \ddots & \vdots \\
      F^l_{ \spin \spin' N0}  & \cdots & F^l_{ \spin \spin' NN} 
    \end{array}\right)
\end{equation}
with the Hartree shift 
\begin{equation}
J^l_{\spin\spin' nn'} = U^\mathrm{H}_{\spin \spin'} \int_0^\infty dr\, r^2 R^l_{\spin n }(r) n_{\spin'}(r) R^{l}_{\spin n'}(r)
\end{equation}
and the pairing field 
\begin{equation}
F^l_{\spin\spin'nn'} = \int_0^\infty dr\, r^2 R^l_{\spin n }(r) \gap (r) R^{l}_{\spin' n'}(r).
\end{equation}
The connection between the interaction strength $U^\mathrm{H}_{\spin \spin'}$ used in the Hartree shift 
and the bare interaction strength $U_{\spin \spin'}$ will be discussed below.
The energy matrix $\mbox{\boldmath$\epsilon$}^{}_{l \spin}$ is diagonal with 
elements $\epsilon^{}_{\spin nl}= \hbar \omega_{\spin}(2 n+l+3/2) - \mu^{}_\spin$ and the operator vectors are $\mb c^{}_{\spin l} = [c^{}_{\spin 0 l 0} \cdots c^{}_{\spin N_l l 0}]^\mt T$. We denote the number of single-particle states with fixed $l$, whose
energy is below the cutoff, as $N_l = \left[ E_\mathrm{c}/(\hbar \omega) - l -3/2\right]/2$.

Similarly to free space, the $H_l$ matrices can be diagonalized using the 
Bogoliubov transformation which is provided by unitary $3 N_\mathrm{c}^l \times 3 N_\mathrm{c}^l$-matrices $\mb W^l$. By inserting the
identity operator $(\mb W^l)^\dagger \mb W^l$ between the matrix and 
the operator vectors, we have the quasiparticle basis as  
$\left(\mb \gamma^{}_{1 l} \, \mb \gamma^\dagger_{2 l} \, \mb \gamma^{}_{3 l}\right)^T = \mb W^{l} \left( \mb c^{}_{1 l} \, \mb c^\dagger_{2 l} \, \mb c^{}_{3 l} \right)^T$.
The rotation matrix $\mb W^{l}$ is chosen such that the matrix in~(\ref{matrixform}) is diagonalized.

The equations for the pairing fields and the densities are
\begin{equation}
\fl
    \Delta_{\spin \spin'}(r) = 
    \tilde U_{\spin \spin'}\sum_{nn'l} (2l+1) R^l_{\spin n}(r) R^l_{\spin' n'}(r) \sum_j W^l_{\bar \spin N_l+n, j}  W^l_{\bar \spin' N_l+n', j} n_\mathrm{F}(E_j),
\end{equation}
and 
\begin{equation}
\fl
n_{\spin}(r) = \sum_{nn'l} (2l+1) R^l_{\spin n}(r) R^{l}_{\spin n'}(r) \sum_j W^l_{\bar \spin N_l+n, j} W^l_{\bar \spin N_l+n', j} n_F((-1)^{\bar \spin} E_j),
\end{equation}
where $n_\mathrm{F}$ is the Fermi distribution and $\bar \spin = \spin -1$, $\spin \in \{1,2,3\}$.
The total number of particles in each component $|\spin\rangle$ is obtained by integration
$N_{\spin} = \int_0^\infty d r \, r^2 n_\spin(r)$.

As in usual BCS theory, the gap equation is ultraviolet divergent, hence the
energy cutoff $E_\mathrm{c}$. In order to make the model cutoff independent,
we follow a standard approach~\cite{Grasso2003a} and use a renormalized
interaction $\tilde U_{\spin \spin'}$ but now 
generalized to a three-component system
\begin{equation}
    \frac{1}{\tilde{U}_{\spin \spin'}(r)} = \frac{1}{U_{\spin \spin'}}-\frac{m_r k_{\mathrm{c},\spin\spin'}(r)}{2 \hbar^2 \pi^2} \alpha_{\spin\spin'}(r)
\end{equation}
where
\begin{equation}
\alpha_{\spin\spin'}(r) = 1 - \frac{1}{2} \frac{k_{\mathrm{F},\spin\spin'}(r)}{k_{\mathrm{c},\spin\spin'} (r)} \ln \ls \frac{k_{\mathrm{c},\spin\spin'} (r) + k_{\mathrm{F},\spin\spin'}(r)}{k_{\mathrm{c},\spin\spin'} (r)-k_{\mathrm{F},\spin\spin'}(r)}\rs.
\end{equation}
Here the momentum cutoff $k_{\mathrm{c},\spin \spin'}$ and local Fermi momentum $k_{\mathrm{F},\spin \spin'}$ are defined as
\begin{equation}
\fl \frac{\hbar^2 k_{\mathrm{c}, \spin \spin'}^2(r)}{2m_r} =  \lh E_c - \bar \mu^{}_{\spin \spin'}(r) \rh \quad \mathrm{and} \quad
  \frac{\hbar^2 k_{\mathrm{F},\spin \spin'}^2(r)}{2m_r} = \lh 2 \bar \mu^{}_{\spin \spin'}(r)- W_\spin(r) -W_{\spin'}(r) \rh,
\end{equation}
where the local average chemical potential is
\begin{equation}
\bar \mu^{}_{\spin \spin'}(r) =  \lh \mu^{}_\spin - V_\spin(r)+ \mu^{}_{\spin'}  -V_{\spin'}(r) \rh/2.
\end{equation}

\begin{figure}
  \centering
  \includegraphics[width=0.70\textwidth]{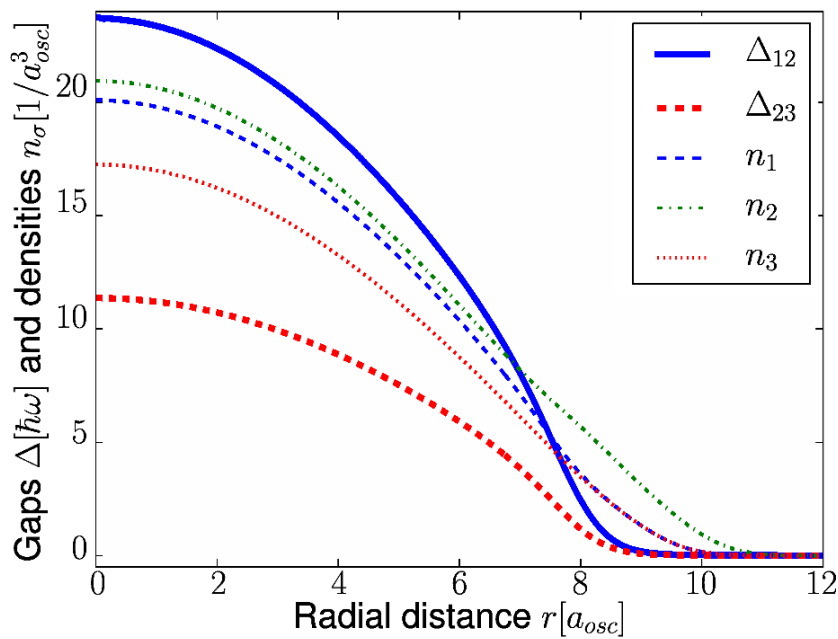} 
  \caption{Typical gap and density profiles in harmonic trap calculated
for temperature $T = 0.05\,T_\mathrm{F}$. Other parameters are the same
as in Figure~\ref{fig:profiles}.}
  \label{fig:temp}
\end{figure}

\begin{figure}
  \centering
  \includegraphics[width=0.70\textwidth]{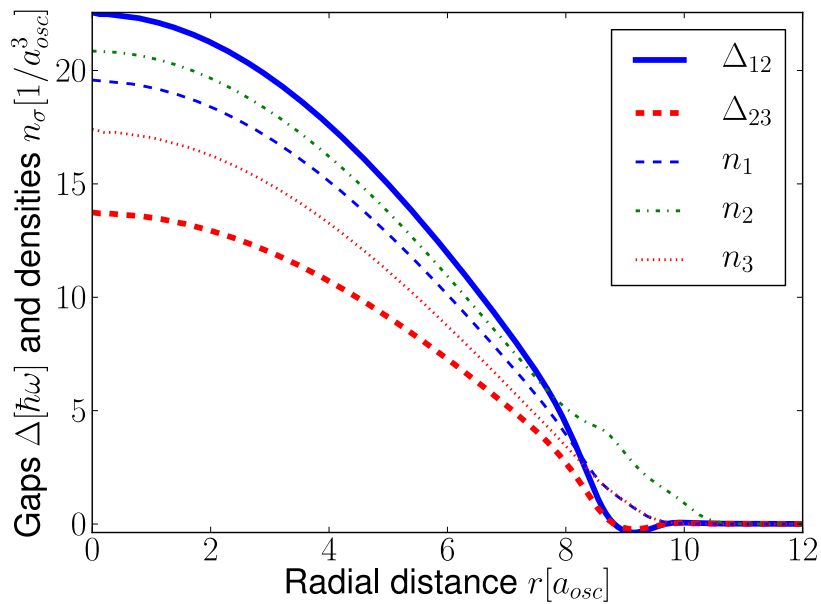} 
  \caption{Typical gap and density profiles in harmonic trap showing a large
coexistence region of the pairing fields $\Delta_{12}$ and $\Delta_{23}$. 
The parameters are $N_2 = 3\times 10^4$, $N_1 = 0.8 N_2$, $N_3 = 0.7 N_2$, and $(k_F a_{12})^{-1} = (k_F a_{23})^{-1} = -0.50$.}
\label{fig:profiles}
\end{figure}

We solve these gap and number equations self-consistently using 
fixed point iteration. For every iteration step in the gap equation, 
we solve the chemical 
potentials $\mu^{}_\sigma$ to keep the particle numbers constants. The 
iteration is terminated when the subsequent gap profiles in the iteration differ 
by at most $5 \times 10^{-5}\, \hbar \omega$.
The cutoff energy is chosen to be $2.5 \times \mbox{max}\{E_\mathrm{F}^\spin\}$ ,
(with a higher cutoff, the results do not qualitatively change). 
If the convergence is slow, we try different initial values to ensure that the 
final result is correct. We use a small finite temperature 
($T = 10^{-3}\,T_\mathrm{F}$) to smoothen the Fermi distribution and to help 
solving the number equations in 
presence of a discrete energy spectrum. However, we have checked that the
results are unchanged even for zero temperature for example in Figures~\ref{fig:profiles} and~\ref{fig:profiles2}. 
We do not consider higher temperatures in this work, however, we have checked 
that our results are sufficiently robust to survive low but experimentally 
relevant temperature $T = 0.05\,T_\mathrm{F}$. An example of the effect of the 
temperature is shown in Figure~\ref{fig:temp} where exactly the 
same parameters
as in Figure~\ref{fig:profiles} are used except that 
$T = 0.05\,T_\mathrm{F}$. The results, especially the coexistence region, 
are practically identical, only the minor features at the edge of the gas have
been smoothened.

The Hartree fields become infinite with a diverging scattering length 
$a_{\spin \spin'} \rightarrow \infty$. This unphysical effect is caused by
improper treatment of two-body scattering effects and in practice these energy
shifts are limited by the Fermi energy. Monte Carlo results on a two-component
Fermi gas suggest that the Hartree fields at unitarity do not exceed
$|W| \approx 0.5\,E_\mathrm{F}$~\cite{Magierski2008a, Haussmann2009a}.
We limit the Hartree field interaction to be smaller than this value by
imposing a hard cutoff on the Hartree interaction strength. That is, instead
of the bare interaction $U_{\spin \spin'}$ we use $U^\mathrm{H}_{\spin \spin'}$
which is limited from above by
\begin{equation}
  \left|U^\mathrm{H}_{\spin \spin'} n_{\spin'} (0)\right| \leq 0.5 E_\mathrm{F}^{\spin'}.
\end{equation}
Notice that the component $|2\rangle$ experiences two Hartree fields due to the
two components $|1\rangle$ and $|3\rangle$ and thus the total Hartree
shift experienced by this component can be up to double the above cutoff.
Since the Hartree shifts induce a mismatch between the  Fermi surfaces, 
the pairing amplitude is reduced. 
Notice that this is in contrast to the balanced two-component case in which 
both components experience the same Hartree potential and the densities 
remain thus equal.  In three component systems, the inclusion of the third
interaction $U_{13}$ would reduce the mismatch, but because it is
usually weaker, the mismatch does not totally disappear.
However, the mismatch can be countered by careful choice of interaction 
strengths and atom numbers, so that local density imbalances are reduced.

\begin{figure}
  \centering
  \includegraphics[width=0.70\textwidth]{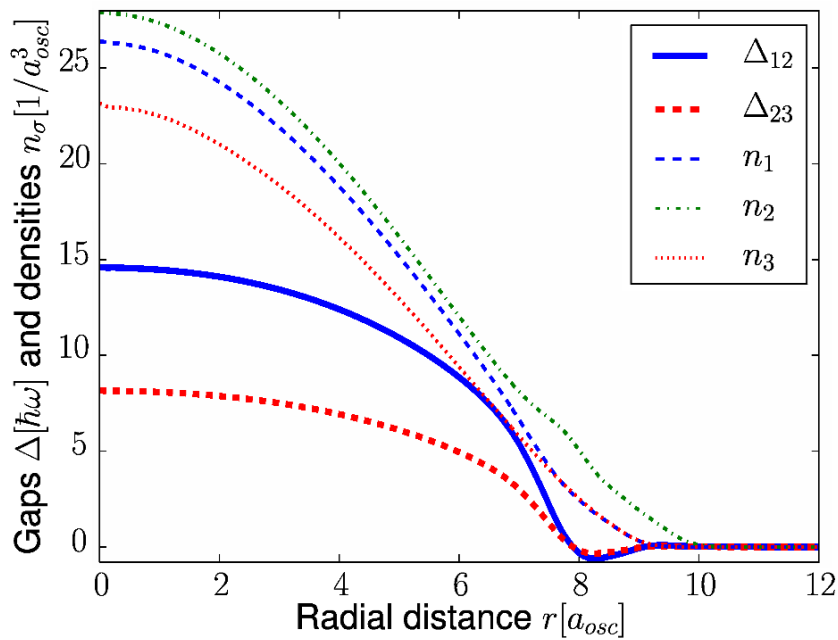} 
  \caption{Typical gap and density profiles in harmonic trap calculated
for zero temperature but including the Hartree energy shift in the way 
described in the main text. The parameters are the same as in Figure~\ref{fig:profiles}.}
\label{fig:hart}
\end{figure}

The Hartree shift can be seen to produce interesting shell structures
for certain parameters. Such exotic shell structures
created by the Hartree shift will be considered elsewhere, requiring a more
complete treatment of the Hartree effect to confirm their validity.
In the present work, we have chosen to focus on a parameter range in which
these peculiar features are not present and in all of the results shown in 
this work except in Figure~\ref{fig:hart} we neglect the Hartree 
effects, i.e. we use $U^\mathrm{H}_{\spin \spin'} = 0$ for all $\spin,\spin'$. 
Our choice does not significantly 
limit the parameter range, because these effects appeared always only in 
tiny islands in the parameter space, typically at the edges of the trap.
Indeed, we have checked that the 
inclusion of the Hartree shift in the way described above does not 
qualitatively change the results presented here. An example is shown in 
Figure~\ref{fig:hart},
which presents the case of Figure~\ref{fig:profiles} but with Hartree fields included. The 
qualitative
behaviour is the same although the numerical values of the order parameters are
smaller. Importantly, the Hartree fields do not affect the coexistence of the 
two order parameters.

\section{Results}
\label{sec:results}

\begin{figure}
  \centering
  \includegraphics[width=0.70\textwidth]{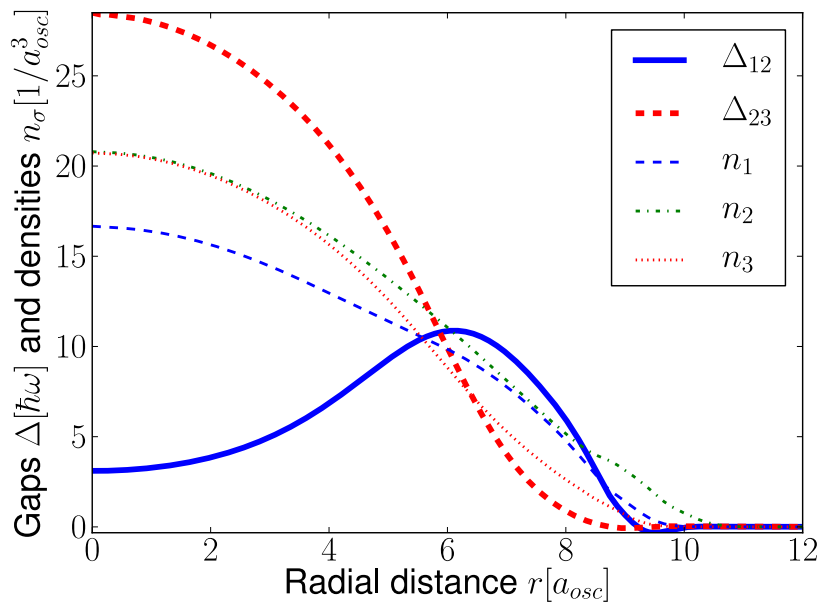}
  \caption{Gap and density profiles as in Figure~\ref{fig:profiles} but
for mismatched interaction strengths $(k_F a_{12})^{-1} = -0.50$, $(k_F a_{23})^{-1} = -0.45$. The two
pairing fields $\Delta_{12}$ and $\Delta_{23}$ separate into different regions,
with a $\Delta_{23}$ core surrounded by a $\Delta_{12}$ shell.}
\label{fig:profiles2}
\end{figure}

Figures~\ref{fig:profiles} and~\ref{fig:profiles2} show typical gap and density profiles obtained from the 
BdG method. In the former the gaps follow the density distributions, and 
the two pairing gaps $\Delta_{12}$ and $\Delta_{23}$ are present across 
the trap. 
In the latter the gaps are spatially separated, with the 
$\Delta_{12}$ pairing field 
concentrated at the edge of the trap and $\Delta_{23}$ at the center of the 
trap.
There is no clear interface between the two pairing regions, but the 
penetration length of pairing field $\Delta_{12}$ inside $\Delta_{23}$ is 
relatively constant when increasing the system size 
(increasing atom numbers $N$), characteristic length scale given by the
oscillator length $r_\mt{osc} = \sqrt{\hbar/m_2 \omega_2}$. Locally the
dominating pairing channel is the one for which the atom densities are 
least mismatched, the strength of the interaction being only a secondary
factor. This local nature of pairing allows interesting shell 
structures~\cite{Paananen2007a} as shown in Figure~\ref{fig:profiles2}.
However, we do not pursue these issues here but rather concentrate
on more general features. 
Notice that the number of atoms in component $|3\rangle$ has been chosen
to be slightly smaller than in components $|1\rangle$ and $|2\rangle$, but 
the features shown here are very general.

\begin{figure}
  \centering
  \includegraphics[width=0.45\textwidth]{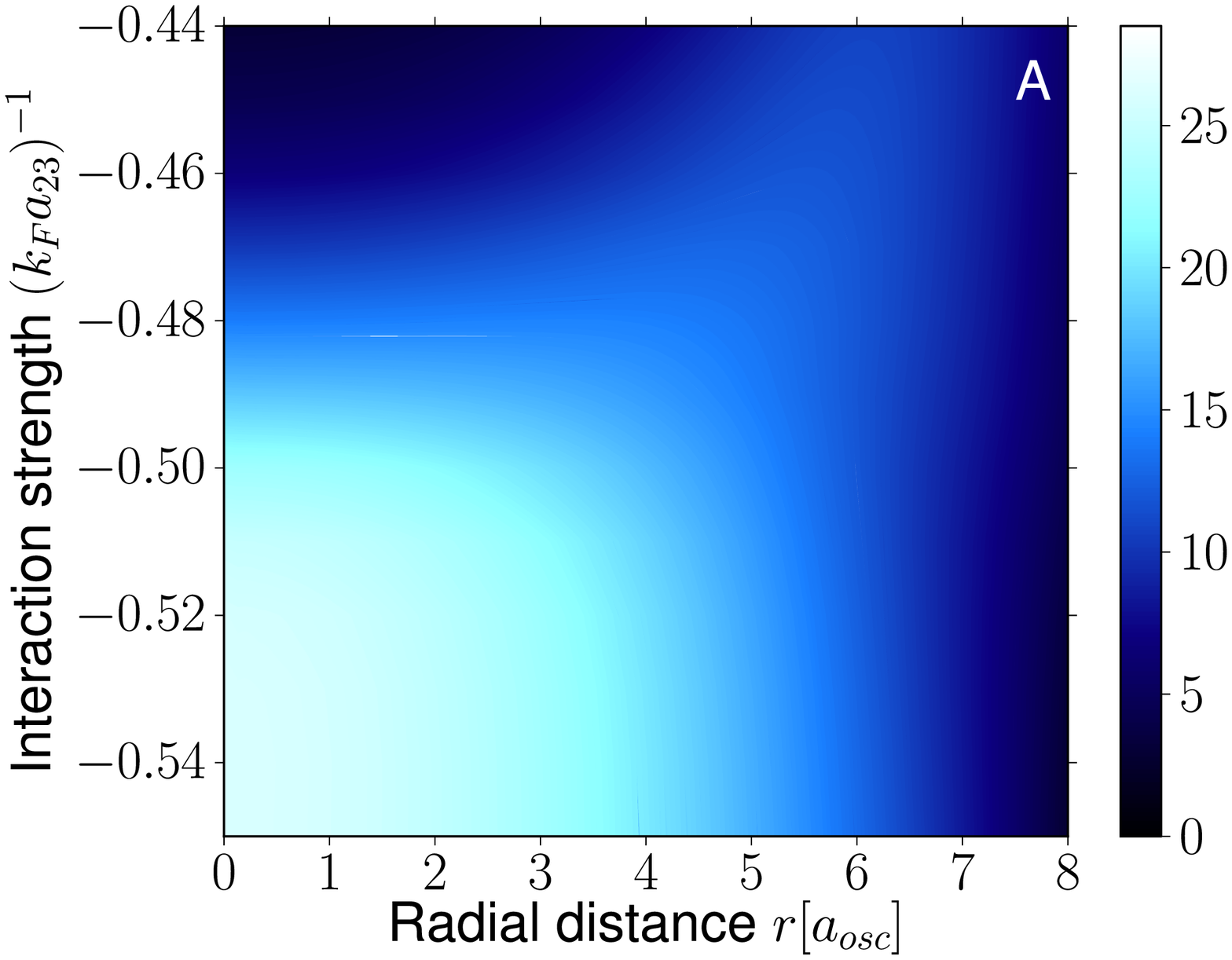} 
  \includegraphics[width=0.45\textwidth]{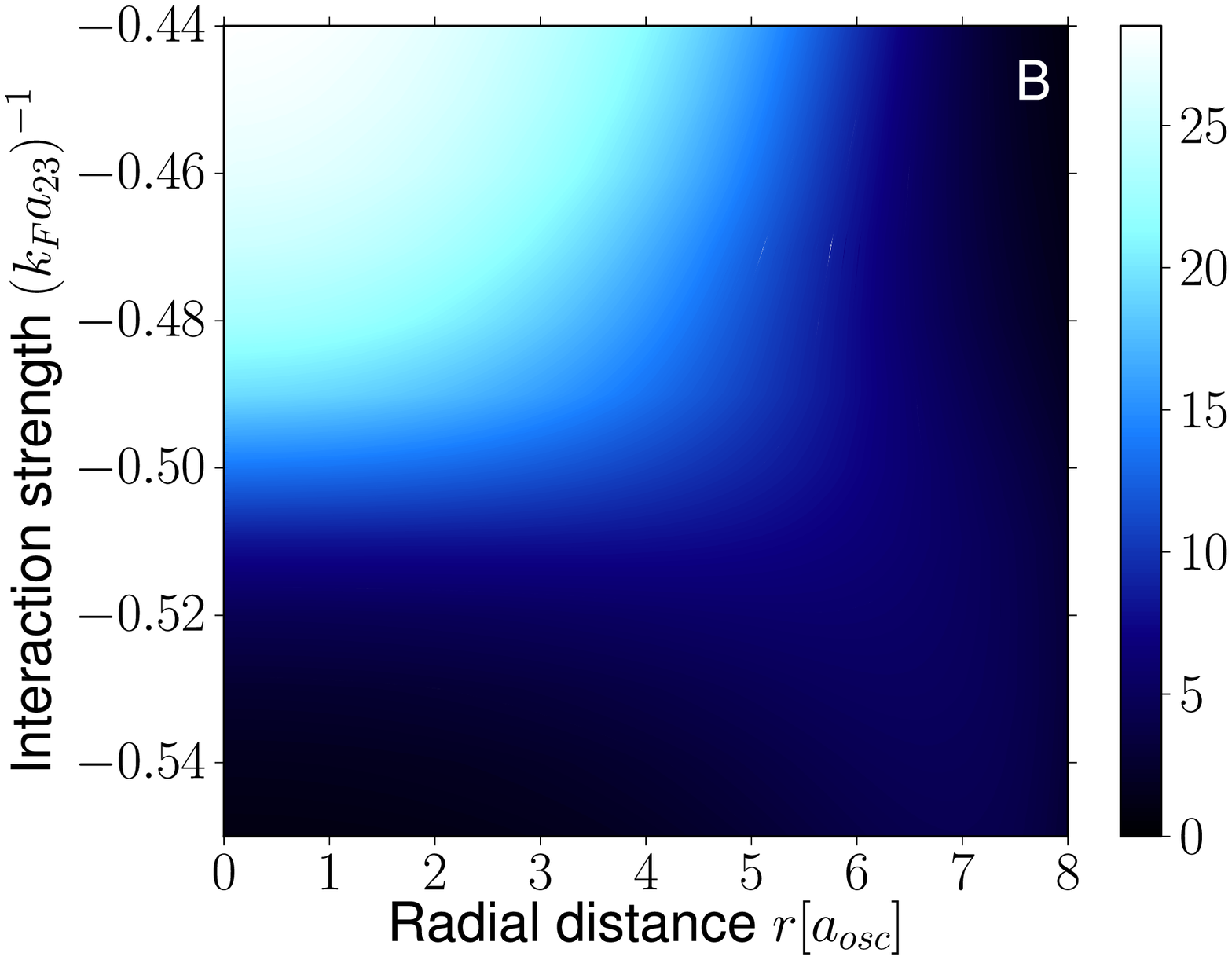}\\
  \includegraphics[width=0.45\textwidth]{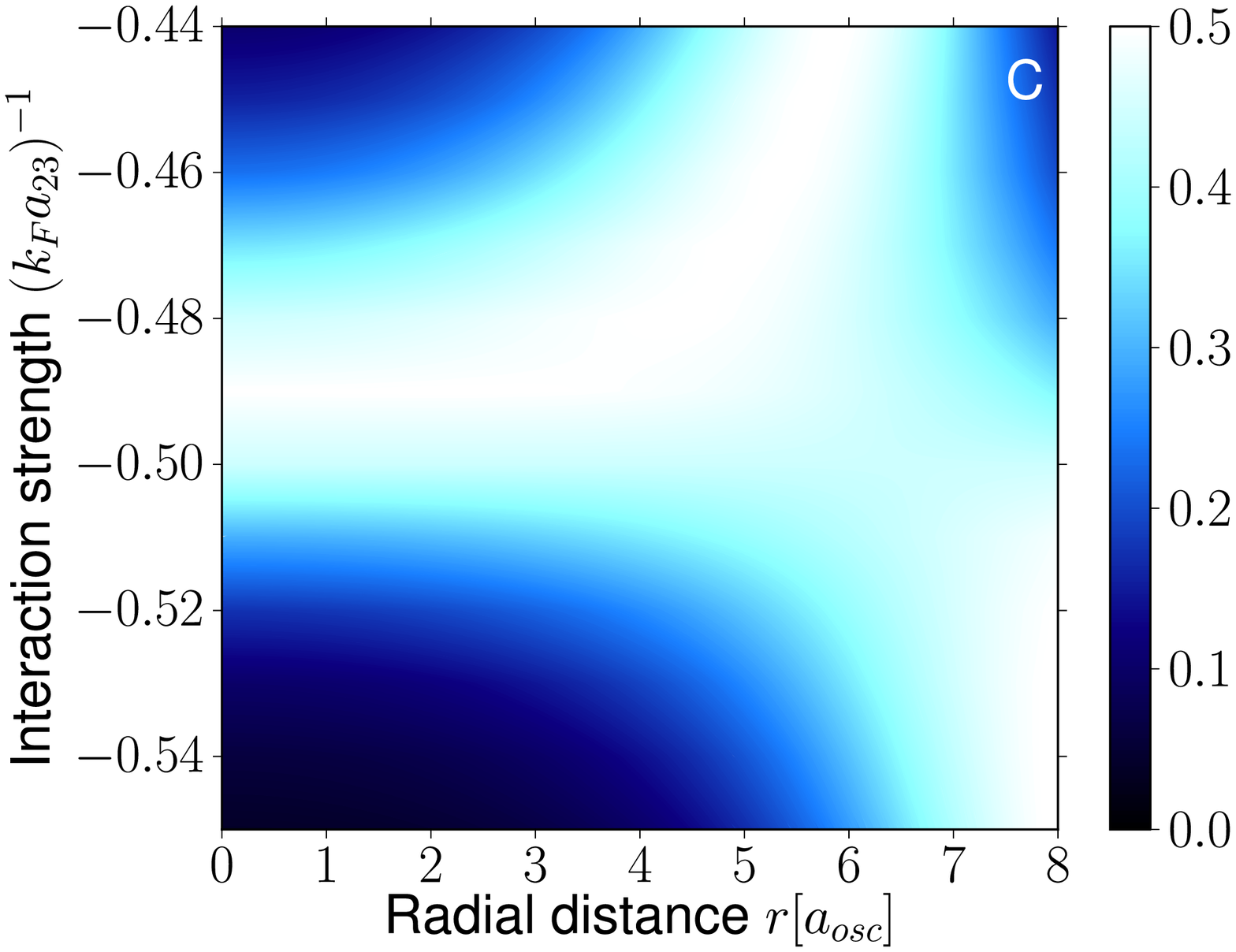} 
  \includegraphics[width=0.45\textwidth]{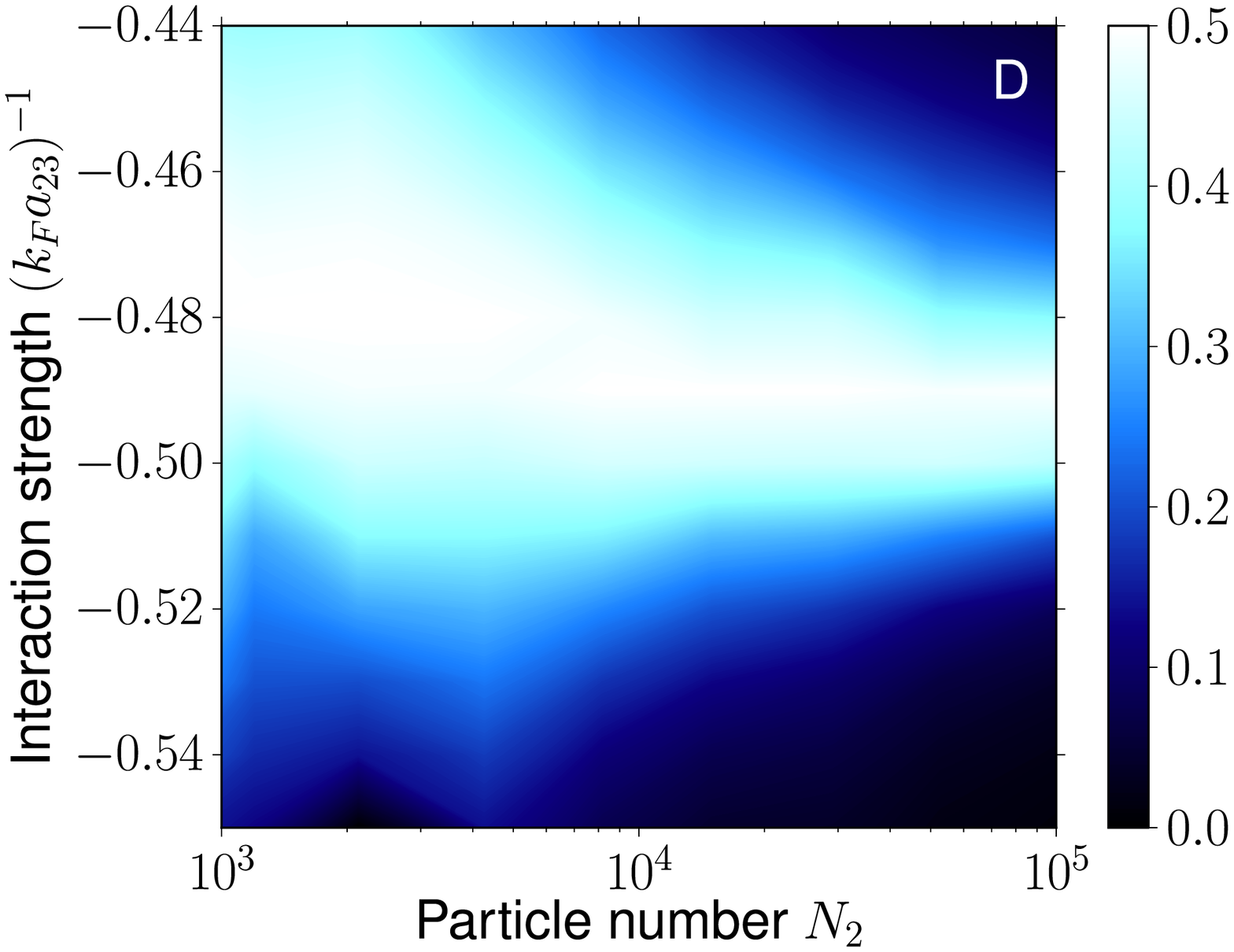}
  \caption{(a) Pairing field $\Delta_{12}$(in units of $\hbar \omega$) as a 
function of the interaction strength $U_{23}$. (b) Pairing field $\Delta_{23}$. 
(c) The coexistence parameter $P_\mathrm{co} = \Delta_{12}\Delta_{23}/(\Delta_{12}^2 + \Delta_{23}^2)$ shows the coexistence areas. Atom numbers in a), b) and c) are the same as in Figure~\ref{fig:profiles}. In (d) the coexistence parameter is plotted at $r = 0$ for different numbers of particles $N$.}
\label{coex1}
\end{figure}

Figures~\ref{coex1} a) and b) show the pairing fields $\Delta_{12}$ and 
$\Delta_{23}$ as a function of interaction strength $(k_F a_{23})^{-1}$.
When either of the two interaction strengths $a_{12}$, $a_{23}$ is 
significantly stronger, the corresponding pairing channel will dominate. 
The crossover between the two regions occurs
at $(k_F a_{23})^{-1} = -0.46$ (not at $(k_F a_{23})^{-1} = -0.50$ because of
the atom number mismatch $N_1 > N_3$). To better characterize the 
coexistence of the two pairing gaps, we define a dimensionless 
\emph{coexistence} parameter
\begin{equation}
  P_\mathrm{co} = \frac{\Delta_{12} \Delta_{23}}{\Delta_{12}^2 + \Delta_{23}^2}.
\end{equation}
Figure~\ref{coex1} c) shows this parameter as a function of interaction 
strength and position, revealing a large coexistence region in the 
somewhat narrow interaction
strength window $-0.51 < (k_F a_{23})^{-1} < -0.47$, but also a coexistence
region close to the edge of the trap across a wide range of interactions.
In Figure~\ref{coex1} d) we show how the coexistence parameter at the 
center of the trap scales with
increasing system size $N$ (the atom numbers $N_1$ and $N_3$ are scaled
correspondingly so that the relative polarizations are fixed).
The coexistence area is suppressed as $N$ grows large, implying that 
the coexistence may vanish in the thermodynamic $N \rightarrow \infty$ limit.
However, with sufficiently accurate choice of interaction strengths, the
coexistence region should be experimentally accessible with reasonably
sized atom gases. We have not studied the scaling of the coexistence regions
at the edge of the trap, but since the penetration length in Figure~\ref{fig:profiles2} is given by the oscillator length we expect the $N\rightarrow \infty$
limit to yield a phase separation into a $\Delta_{23}$ core and a surrounding 
$\Delta_{12}$ shell.

\begin{figure}
  \centering
  \includegraphics[width=0.70\textwidth]{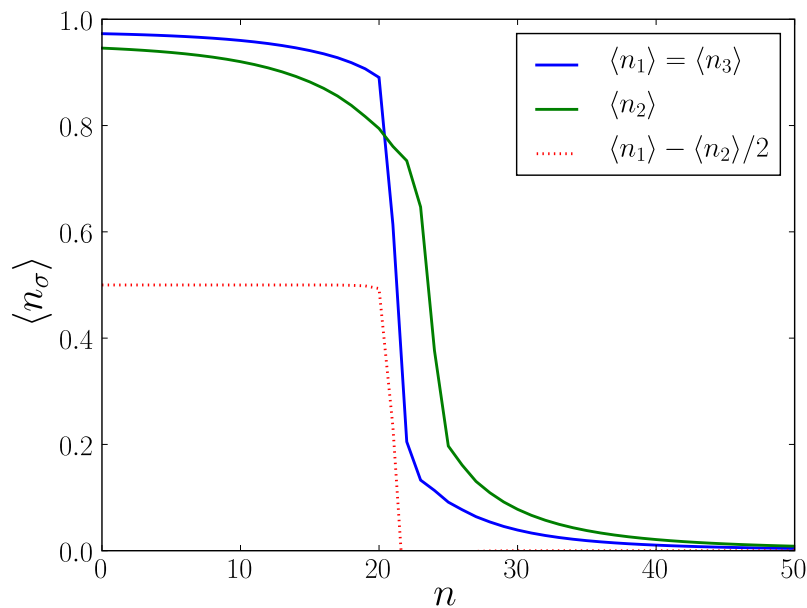}
  \caption{Occupation numbers $\langle n_{\spin n,l=0} \rangle = \langle c_{\spin n 0}^\dagger c_{\spin n 0} \rangle$ for each component $\spin$ as a function
of the quantum number $n$ with $l=0$. The parameters have been chosen 
symmetrically $N_1 = N_3 = 16000$, $N_2 = 20000$, and $\left(k_\mathrm{F} a_{12}\right)^{-1} = \left(k_\mathrm{F} a_{23}\right)^{-1} = -0.50$, resulting in
identical pairing gaps $\Delta_{12}(r) = \Delta_{23}(r)$.}
\label{fig:occu}
\end{figure}

To better understand the nature of the pairing scheme in the coexistence 
areas, Figure~\ref{fig:occu} a) shows the
occupation numbers of different $n$-quantum number states  
for a symmetric case $\Delta_{12}(r) \equiv \Delta_{13}(r)$. The angular 
momentum quantum number $l=0$ chosen here acts as a representative of a 
more general behavior for general $l$.
The figure reveals a very similar pairing scheme as in the homogeneous system, see
Figure~\ref{fig:free_momentum}. The unpaired atoms are distributed among
the two components $|1 \rangle$ and $|3 \rangle$ and form a step at
the Fermi surface.

\section{Conclusions}
\label{sec:conclusions}

We have studied the pairing of a three-component Fermi gas 
when two of the interspecies interaction channels are dominant. In 
a homogeneous system, we showed that different boundary conditions, 
namely fixing the chemical potential or fixing the atom numbers, produce 
qualitatively different results and phases. We have not considered the 
possibility of a phase separation in which the two pairing fields would
be spatially separated in otherwise uniform system. 

For trapped systems, our BdG study reveals an interesting coexistence 
region where both pairing channels $\Delta_{12}$ and $\Delta_{23}$ are 
present. This is a mesoscopic effect and likely to vanish in the
limit of a large system $N \rightarrow \infty$ resulting in phase separation
into shells of different pairing fields. However, the coexistence region
is present at atom numbers relevant for atom gas experiments, making the observation
of this intriguing double-gap prediction feasible.

There is already a wide range of standard experimental techniques to detect
such pairing correlations. For example, one could use 
radio-frequency spectroscopy~\cite{Torma2000a,Chin2004a,Kinnunen2004a}
for driving atoms from the hyperfine states $|1\rangle$ and $|3\rangle$ separately into
some fourth noninteracting state $|e\rangle$. Other possibilities include transforming
pairing correlations into molecular $12$ and/or $23$ pairs through magnetic field 
sweeps~\cite{Regal2004b} or optical molecular spectroscopy~\cite{Partridge2005a}.

\section*{Acknowledgements}

We acknowledge funding from Academy of Finland and EUROQUAM/FerMix (Project No. 210953, No. 213362, No. 217045, No. 217043). 
This work was conducted as a part of a EURYI scheme grant, see www.esf.org/euryi.

\section*{References}

\bibliographystyle{unsrt}
\bibliography{ref}

\end{document}